\documentclass[floatfix,showkeys,twocolumn,showpacs,preprintnumbers]{revtex4}
\usepackage{graphicx}
\usepackage{amssymb}

\begin{document}

\title{Predicting the species abundance distribution using a model food web}

\author{Craig R. Powell}
\email{craig.powell@manchester.ac.uk}
\affiliation{Theoretical Physics Group, School of Physics and Astronomy, University of Manchester, Manchester, M13 9PL, UK}

\author{Alan J. McKane}
\email{alan.mckane@manchester.ac.uk}
\affiliation{Theoretical Physics Group, School of Physics and Astronomy, University of Manchester, Manchester, M13 9PL, UK}

\date{\today}

\begin{abstract}
A large number of models of the species abundance distribution (SAD)
have been proposed, many of which are generically similar to the
log-normal distribution, from which they are often indistinguishable
when describing a given data set.  Ecological data sets are
necessarily incomplete samples of an ecosystem, subject to
statistical noise, and cannot readily be combined to yield a closer
approximation to the underlying distribution.  In this paper we use
empirical data obtained from an ecosystem model to study the predicted
SAD in detail, resolving features which can distinguish between models
but which are poorly seen in field data.  We find that the power-law
normal distribution is superior to both the log-normal and
logit-normal distributions, and that the data can improve on even this
at the high-population cut-off.
\end{abstract}



\maketitle

\section{Introduction}
\label{intro}
The species abundance distribution (SAD) is one of the most widely
studied descriptions of an ecological community.  To determine it, the
number of species in a given community which have $n$ observed
individuals is plotted against $n$.  The shape of this plot has been
investigated by a great many empiricists and theorists over the years,
beginning with the classic work of \citet{fis43} and \citet{pre48}.
Reviews of the subject \citep{whi65,may75,gra87,mar03,mcg07} reveal
the large number of mechanisms that have been proposed to explain the
observed SAD.  Many of these mechanisms predict the essential aspects
of the observations, that is, a few very abundant species and many
rare species.  As a consequence it has become very difficult to
falsify proposed mechanisms from empirical data, which has led to the
authors of the most recent multi-author review \citep{mcg07} to
contrast the development of the analysis of SADs with ``a healthy
scientific field'' in which ``theoretical, empirical and statistical
developments [...] advance roughly in parallel''.

In this paper we suggest a way forward which is in effect intermediate
between the theoretical and empirical approaches.  We measure the SAD
in an established model which constructs an ecological community as
a set of predator-prey interactions \citep{dro01}.  The model itself
was originally created so that many of its key properties were
emergent and not put in by hand.  So, for instance, trophic levels
emerge from the nature of the predator-prey interactions; species are
not labelled as ``plants'', ``herbivores'' or ``carnivores'' a
priori.  This contrasts with traditional theoretical approaches which
either postulate a mechanism, or if a model community is put forward
it is usually rather simple, with the form of the SAD following from
one of the fundamental aspects of the theory.  Conversely,
measurements taken in the field will of necessity include numerous
influences involving climate, terrain, location etc., which are not
present in the model we use to measure SADs.  Thus the SADs we measure
will be free of these external influences, but still be determined by
influences which are too complex to easily characterise.  This
approach will also allow us to measure SADs for a multi-trophic level
community whereas, so far as we are aware, all other predictions for
SADs have been for communities of trophically similar species.

The model we will be using (called the Webworld model) has been
developed over a number of years \citep{cal98,dro01,dro04,qui05}.  In
it, species are defined in terms of traits (phenotypic and behavioural
characteristics), and it is the nature of the interactions between
these traits which define the nature of interactions between species.
This community is built up from a small number of species through a
speciation mechanism which creates a new species with a novel set of
features.  Resources are distributed through a quite elaborate set of
equations governing population dynamics with adaptive foraging.
Overviews of the model are given in review articles
\citep{dro03,mck05,mck06}, and more briefly in section~\ref{Sec:model}.
In  section 3 we outline the method of our analysis, in
section 4 we describe the results obtained and we conclude with a review
of the results in section 5.

\section{Model description}
\label{Sec:model}
The Webworld model consists of a set of species, each defined by its
unique combination of ten different features.  The features are chosen
from a set of $L$ possible features determined at the start of the
simulation, at which point two species are created.  One of these is
the environment species, which has a fixed population for all time and
is the ultimate source of energy for all species in the food web.  The
other initial species is the common ancestor of all species encountered
during a simulation run.

The dynamics of the Webworld model occur on three separated
time-scales.  The longest of these is the evolutionary time-scale, on
which new species are added as mutated versions of extant species.
Specifically to the implementation of Webworld, a species  is selected
at random without regard to its population, except that this must be
non-zero.  One individual of that species is then used to found a new
species identity, sharing all features but one with the parent
species.  The remaining feature is selected to avoid repetition either
of the same feature within one species or the same set of features
between species, but is otherwise selected at random.  The newly introduced
species is then subject to the same population dynamics as all other
species, which is the dynamical process that occurs on the
intermediate time-scale.

Population dynamics occurs by balance of energy; energy is gained
through ``predation'', which in the case of feeding on the environment
species we interpret as autotrophy.  Each species $i$ changes its
population $n_i$ according to the balance equation
\begin{equation}
  \dot n_i=\lambda\sum_jg_{ij}n_j - \sum_jg_{ji}n_i - dn_i,
  \label{eq:popdyn}
\end{equation}
where $g_{ij}$ is the functional response, the dependence of the rate
of energy consumption by species $i$ on the population of species
$j$.  The factor of $\lambda=0.1$ introduces an ecological efficiency
whereby the energy lost to species $j$ is greater than that gained by
its predator $i$.  Thus the first term on the right hand side of
Eq.~(\ref{eq:popdyn}) is the energy income of species $i$ summed
across all prey species, while the second term is the energy loss
summed across all predators.  If species $a$ does not feed on species
$b$ then $g_{ab}=0$, and hence this does not contribute to either
sum.  The final term in Eq.~(\ref{eq:popdyn}) is the loss of energy
from species $i$ due to death of its constituent individuals at rate
$d$ per individual; the expected lifespan of an individual is
therefore $1/d$, which for simplicity we take to be the same across
all species.  Death appears in our model purely as an energy
loss term and cannot be made an evolvable quantity, since it has a
preferred value of zero.

The shortest time-scale in the Webworld model reflects the choice of
foraging strategy by individuals of each species.  The functional
response for Eq.~(\ref{eq:popdyn}) is given by
\begin{equation}
  g_{ij} = \frac{f_{ij}S_{ij}}{bn_j+\sum_k\alpha_{ik}f_{kj}S_{kj}n_k}n_i,
  \label{eq:functionalResponse}
\end{equation}
where $f_{ij}$ is the fraction of its time species $i$ spends feeding
on prey species $j$, which is the quantity to be optimised in order to
maximise $\sum_jg_{ij}n_j$.  $S_{ij}$ and $\alpha_{ij}$ are constants
defined by relating the features of species $i$ and $j$, $S$
indicating the ability of $i$ to feed on $j$, and $\alpha$ relating to
the degree of inter-specific competition.  To prohibit mutual
predation the matrix $S$ is made anti-symmetric, thus
$S_{ij}=-S_{ji}$, and the shortest possible feeding loop involves
three species.  Matrix $\alpha$ is symmetric, with maximum competition
$\alpha_{ii}=1$ between members of the same species, and minimum
competition 0.5 between highly-different species.  By calculating $S$
and $\alpha$ based on a set of features largely conserved during
speciation we ensure that each newly-founded species has similar
abilities to its parent species, with which it is also in strong
competition, and in particular the dynamics of two identical species,
were they allowed, would be indistinguishable from the dynamics of
pooling them as one species. 

In \citet{dro01} an evolutionarily stable strategy was shown to exist 
for foraging, which can be found by iteratively solving
Eq.~(\ref{eq:functionalResponse}) with the condition that
\begin{equation}
  f_{ij}=\frac{g_{ij}}{\sum_kg_{ik}}.
\end{equation}

The result of the repeated application of these dynamics is the
gradual construction of a complex food web.  Species are removed if
their population falls below 1, and the fixed population of the
environment species, $R$, as such determines the expected number of
species present in the food web at any time, though there is a
continual turnover of species and consequent fluctuation in any given
food web measure.  After running the model for a large number of
evolutionary time steps, there is no systematic change in quantities
such as the number of concurrent species, and the food web structure
appears to have matured.  It is on such webs that we examine the
species abundance distribution.

\section{Method}
Using the Webworld model discussed in the previous section we generate
sets of communities for which the ensemble species abundance
distribution (SAD) can be examined in detail.  Because we use the same
set of possible features and the same environment species in each
case, we assume that the underlying SAD does not alter between model
realisations.  In this case it is possible to pool the resultant
communities in order to determine the SAD with improved statistical
noise.  Details of the computational approach are given in
section~\ref{Sec:Computational}.  In section~\ref{Sec:Fitting} we
discuss the functions which we fit to the data, and the optimisation
criteria of the fitting.  In section~\ref{Sec:Generalisation} we
discuss the problems of generalising fits to include communities
differing in size or trophic level.

\subsection{Comparative models}
\label{Sec:Computational}
Although the Webworld model can simulate ecological communities in
reasonable time, to create large complex communities takes
considerable computation, and to generate enough simulations to get
good statistics across a broad range of parameter space is difficult.
We therefore perform the first examination on a variant of Webworld in
which all species feed exclusively on the environment.  Because all
species are basal, the relative populations are determined by the
relative ability, $S$, and competition, $\alpha$, terms between
existing species, which are selected by evolution in the same way as
in the full model.  By avoiding a large part of the computational
effort we are able to generate large numbers of webs for comparison,
and in the results presented here gather statistics from a set of one
hundred model runs for each value of resources, $R$.  In
section~\ref{Sec:Results} we focus on the fitting of food webs with
resources $10^3$, $10^4$, $10^5$ and $10^6$, but simulations were
performed for numerous other values of $R$ within this range to show
that interpolation of the results is reasonable.  The minimum value of
$R$ results in communities with few species, which become
correspondingly harder to characterise in terms of an SAD.  Larger
values of $R$ become increasingly computationally expensive.  Rather
than attempting to extend the range of $R$ to larger values, we
created a total of 900 basal communities at $R=10^6$ for more detailed
analysis of the tails of the distribution.  Because the common
theoretical SADs have been selected based on reproduction of the modal
peak, and are poorly constrained by observations, the tails offer the
largest differences between candidate SADs.  Due to the much larger
computational complexity of the full Webworld model, we have only a
sample of ten comparable food webs for large $R$ from which to deduce
trophic SADs.

\subsection{Fitting method}
\label{Sec:Fitting}
\begin{figure}
  \centering
  \includegraphics[width=0.45\textwidth]{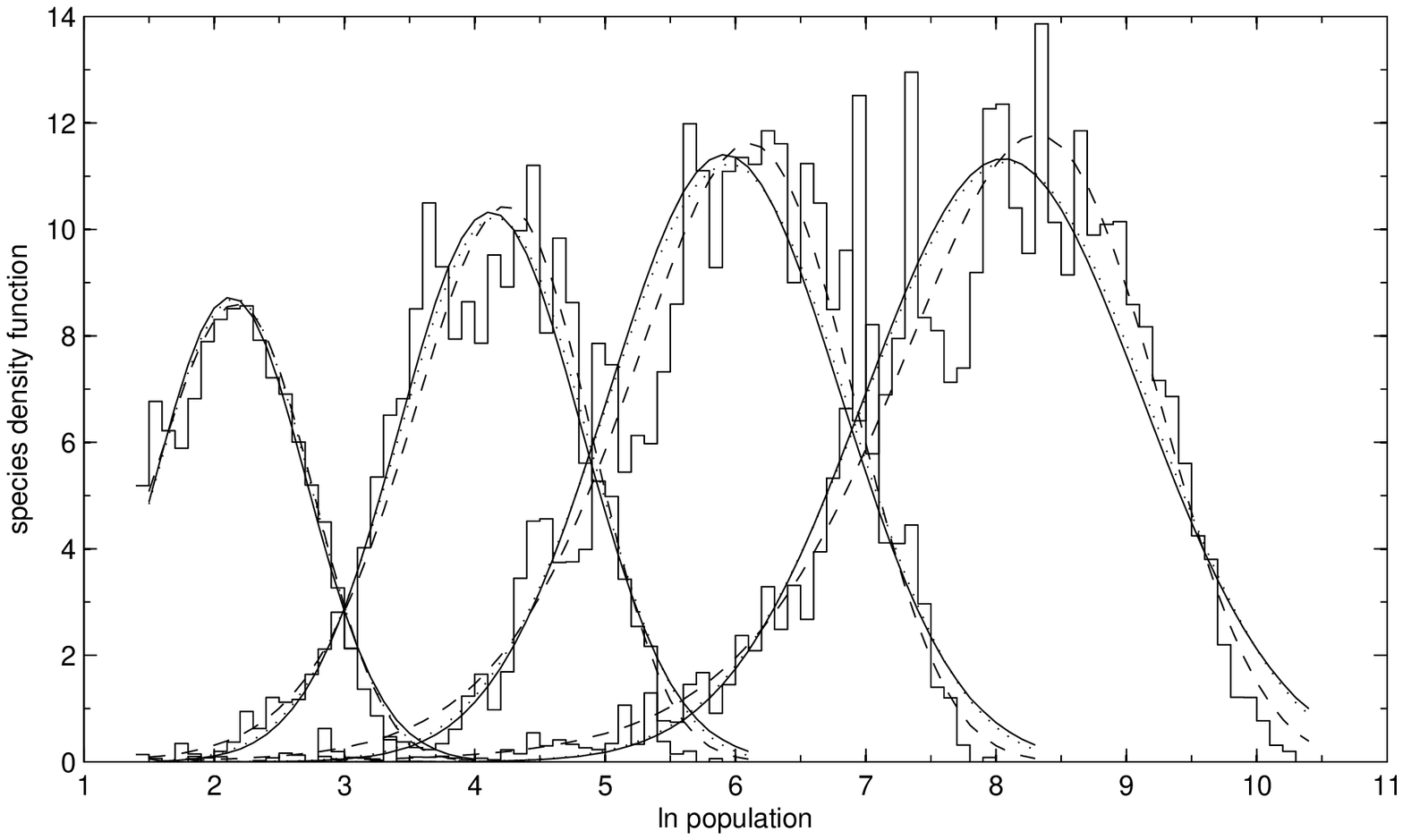}
  \caption{The fitted species abundance distribution for basal
    communities with resources $R=1000$, 10\,000, 100\,000 and
    1\,000\,000. The histogram indicates the data in bins of width 0.1
    in $\ln N$.  The solid curves indicate optimal log-normal fits,
    the dotted lines optimal logit-normal fits, and the dashed lines
    optimal power-law normal fits.  Distributions to the right
    correspond to increasing $R$.
  }
  \label{fig:basalPDF}
\end{figure}

As can be seen in Figure~\ref{fig:basalPDF}, the probability
distribution function (p.d.f.) of species abundance has a rather noisy
histogram even for the largest collection of independent communities
we were able to assemble with the available computer time.  Fitting a
distribution function to such histograms is problematic for several
reasons.  The noise makes it difficult to algorithmically optimise the
fitting function, and hence can obscure differences in the strength of
different functional forms.  More importantly, the apparently optimal
parameters and associated fitness will depend on the arbitrary choice
of bin width and position, since changing these parameters can
significantly alter the distribution of noise between the bins.
Furthermore, the distribution function underlying the observed SAD is
likely to have a functional form other than our approximations, and in
general may be significantly more complicated than we can extract from
data so long as the noise remains.  Rather than obtaining a function
which closely matches the value of the p.d.f. for most population
sizes, but which omits important features, we prefer to recover a
smoothed version of the distribution function which correctly predicts
the total number of species.  As a consequence of these considerations
we fit the integrated version of the fitting function to the empirical
cumulative distribution function (c.d.f.), whose value at a given
population $N$ is the measured number of species with $n_i<N$.  This
definition matches the type of p.d.f. used by \cite{may75} whose
integral is the expected number of species.  P.d.f.s may also be
defined such that the area enclosed is unity.  To illustrate the
fitting procedure we present plots of the measured and fitted c.d.f.s
in addition to the p.d.f.s, and indicate the goodness-of-fit by
plotting the residuals of the c.d.f., that is, the difference between
the integrated fitting function and the measured c.d.f.

The strongest condition that we impose on each fitting function is
that it should correctly predict the number of species more abundant
than the least abundant species observed.  Below this population the
distribution may be terminated by a veil line, but we do not allow any
such consideration for populations above the most abundant species
observed.  Subject to this condition we optimise the parameters of
each theoretical distribution function, $f\!\left(\ln N\right)$, by
minimising a quantity analogous to $\chi^2$.  One such statistic is
the Cram\'er-von Mises test \citep{bai91}, defined as
\begin{equation}
 CM=\frac1{12S}+\frac1{S}\sum_{i=1}^S\left(i-0.5-\hat f\!\left(n_i\right)
\right)^2,
\end{equation}
where $\hat f\!\left(n_i\right)$ is the predicted number of species
less abundant than $n_i$, subject to the veil line at $n_1$, and $S$
is the number of species observed.  Although this is readily
generalised to an ensemble of SADs, it attributes most weight to the
peak of the distribution at the expense of fitting the tails, and we
instead minimise the quantity
\begin{equation}
  k^2=\int_{\ln n_1}^{\ln N_{\rm max}}\left(
  C\!\left(N\right)-\hat f\!\left(N\right)
  \right)^2{\rm d}\ln N,
\end{equation}
where $C\!\left(N\right)$ is the observed number of species less
abundant than $N$.  For many distributions $N_{\rm
max}\rightarrow\infty$, but functions such as the logit-normal
distribution are parametrised by the total number of individuals
observed, $J$, in which case $N_{\rm max}=J$.  Unlike the Cram\'er-von
Mises statistic, $k^2$ places equal weight in all intervals of $\ln
N$.  Given that the theoretical distribution almost certainly differs
from the distribution underlying the data, this tends to avoid
problematic regions, such as ranges of $N$ in which few species are
observed, but where the empirical and theoretical c.d.f.s differ.  The
tails of the distribution often behave in this manner.  Having
identified optimal fitting parameters by minimising $k^2$, we follow
the advice of \cite{mcg07} that ``claim[s] of a superior fit must be
robust by being superior on multiple measures'' by evaluating the
Kolmogorov-Smirnov statistic \citep{hay02} for each theoretical
distribution.  Defined for a single realisation as
\begin{equation}
  d=S^{-1/2}{\rm max}_i\left\{\left|i-1-\hat f\!\left(n_i\right)\right|,
                              \left|i-\hat f\!\left(n_i\right)\right|\right\},
  \label{Eq:KS}
\end{equation}
$d$ corresponds to the greatest deviation between the empirical and
theoretical c.d.f.s.  This must occur at one of the observed species,
which correspond to steps in the empirical c.d.f.  It is necessary to
evaluate the difference between the empirical and theoretical
c.d.f. both immediately before and after the step, and hence the
`maximum' operator in Eq.~(\ref{Eq:KS}) contains two terms for each
observation $i$.  Although the values of $d$ obtained imply rejection
of the theoretical distributions given the amount of data available,
we use $d$ as a measure of the relative goodness-of-fit to distinguish
between theoretical distributions.  Other measures of goodness-of-fit
tend to relate to binned data rather than the c.d.f., and provide
correspondingly weaker evidence \citep{mcg03}.

Although the log-normal distribution has been criticised as
inappropriate for application to SADs \citep{wil05}, it is a commonly
examined form of the SAD and we therefore adopt it as one of the
theoretical SADs we fit to the data.  We also consider the
logit-normal distribution preferred by \citet{wil05}.  Whereas the
log-normal distribution appears as a normal distribution when plotted
against a logarithmic population-axis, the logit-normal has a normal
distribution when plotted against a logit population axis.  Our
analysis will consistently use the logarithmic axis both for plotting
and for the integration of $k^2$, so while the log-normal distribution
has the form
\begin{equation}
  P\!\left(\ln N\right){\rm d}\ln N=A\exp\left\{-\frac{\left(\ln
  N-\ln\mu\right)^2}{2\sigma^2}\right\}{\rm d}\ln N,
\end{equation}
with the fitting parameters $A$, $\mu$ and $\sigma$, the logit-normal
distribution includes an extra factor, giving
\begin{equation}
 P\!\left(\ln N\right)=A\frac{J}{J-N}
\exp\left\{-\frac{\left(\ln\frac{N}{J-N}
 -\ln\frac{\mu}{J-\mu}\right)^2}{2\sigma^2}\right\}.
\end{equation}
We also consider a third fitting function, the power-law normal
distribution, which appears normal against a power-law population
axis.  Transformed to a logarithmic axis, this has the functional form
\begin{equation}
P\!\left(\ln N\right)=A\alpha
N^\alpha\exp\left\{-\frac{\left(N^\alpha-\mu^\alpha\right)^2}
{2\sigma^2}\right\},
\label{Eq:PowerLawNormaldlnN}
\end{equation}
where $\alpha$ is the power-law index.  We do not consider the
log-series distribution since our data are with few exceptions peaked
at large $N$, whereas the p.d.f. of the log-series distribution
decreases from $N=1$ even when drawn against a logarithmic
population-axis.  The broken stick distribution \citep{mag88} was found to be
similar in form to the observed distribution, but inferior to the
log-normal in all cases.

\subsection{Comparison of food webs}
\label{Sec:Generalisation}
\begin{figure}
  \centering
  \includegraphics[width=0.45\textwidth]{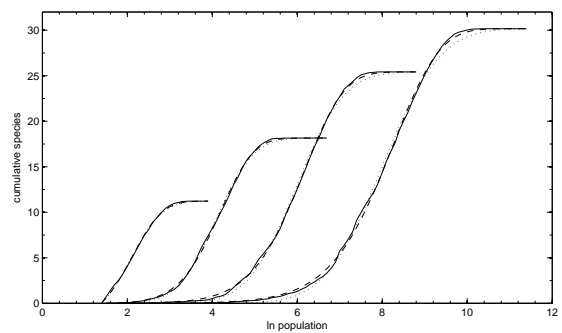}
  \caption{The fitted cumulative species abundance distribution for
    basal communities with resources $R=1000$, 10\,000, 100\,000 and
    1\,000\,000.  The solid line shows the data, the dotted line marks
    the log-normal distribution, and the dashed line the power-law
    normal distribution.
    Distributions to the right correspond to increasing $R$.
  }
  \label{fig:basalCDF}
\end{figure}
Since we are applying the same distribution function with different
parameters to basal food webs of different sizes, and to the SADs of
different trophic levels within a single community, in the ideal case
a parametrisation of the fitting coefficients in terms of resources,
$R$, and trophic level, $l$, would be found.  Because small values of
$R$ correspond to food webs with fewer species, complications arise in
weighting the contribution to goodness-of-fit from differently sized
webs, and we do not in this paper attempt to simultaneously fit webs
of different sizes.  By examination of the best-fitting parameters for
each web we can determine the dependence of parameters on $R$ except
in one case; the power-law index $\alpha$ of the power-law normal
distribution.  For most values of $R$ the goodness-of-fit depends
quite weakly on this parameter, and the optimal value of $\alpha$ is
poorly constrained for any one web.  Since we were unable to identify
a systematic trend or strongly constrain the value of $\alpha$, we
chose $\alpha=0.2$ as a constant value consistent with the optimised
parameters, and fixed this value for all results presented here.

\section{Results}
\label{Sec:Results}
In section~\ref{Sec:Basal} we present the results of the fitting
procedure for the basal communities.  These should give the least
complicated species abundance distributions (SADs), since all species
feed on a single resource and are in direct competition with each
other.  In comparison, the trophic communities examined in
section~\ref{Sec:Trophic} feed on multiple food sources themselves
distributed in abundance, and compete with different subsets of the
other species.  In section~\ref{Sec:Tails} we make use of the large
number of simulation runs which can be performed to make a detailed
examination of the low- and high-population tails of the empirical
distribution, and compare this to the behaviour of the fitted
distributions.

\subsection{Basal community}
\label{Sec:Basal}
\begin{figure}
  \centering
  \includegraphics[width=0.45\textwidth]{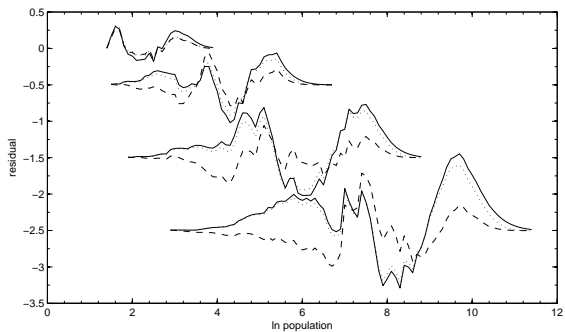}
  \caption{The same data plotted in Figure~\ref{fig:basalCDF} shown as
    residuals; the solid line corresponds to the empirical c.d.f. minus the
    log-normal distribution, the dotted line to the data minus the
    logit-normal distribution, and the dashed line to the data minus the
    power-law normal distribution.  Offsets of -0.5, -1.5 and -2.5 have been
    applied to data for resources $R=10\,000$, 100\,000 and
    1\,000\,000 respectively.
  }
  \label{fig:basalCDFr}
\end{figure}
\begin{figure}
  \centering
  \includegraphics[width=0.45\textwidth]{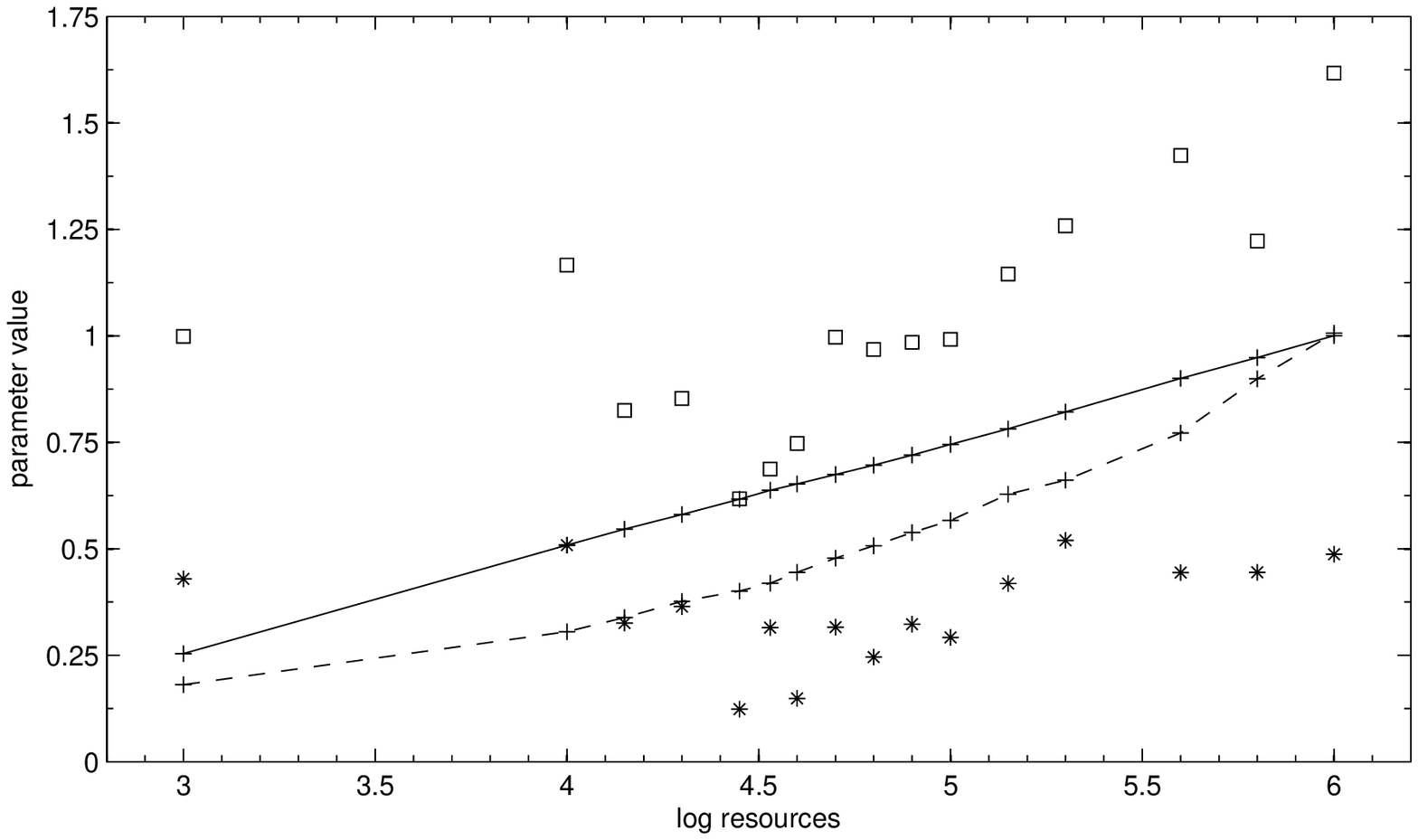}
  \caption{Parameters of the power-law normal fit to the basal
    community SAD for all values of resources examined.  The solid
    line passes through points marking the mean population, $\mu$ in
    Eq.~(\ref{Eq:PowerLawNormaldlnN}); the dashed line marks the
    standard deviation, $\sigma$.  Squares mark the Kolmogorov-Smirnov
    $d$ value, and stars mark the quantity $K$ described in the text.
  }
  \label{fig:basalFitPar}
\end{figure}
The results for this version of the model are the most complete in
that one hundred simulations runs were examined for each value of
resources $R$, and a large number of values of the continuous
parameter were examined.  In Figures~\ref{fig:basalPDF} to
\ref{fig:basalCDFr} only four of these realisations are plotted,
corresponding to $R=10^3$, $10^4$, $10^5$ and $10^6$, which include
the two most extreme values of $R$ for which webs were calculated.
The general features of the SAD for these four values are typical, as
is the goodness of fit achieved by each of the three fitting functions
examined.  It is clear from Figure~\ref{fig:basalPDF} that the
observed distribution is left-skewed (has an over-abundance of rarer
species), a characteristic absent from the log-normal distribution.
The logit-normal distribution does not have significantly improved
skew over the log-normal distribution, since the most abundant species
from any run has less than one quarter of the mean number of
individuals $J$, and the logit function is therefore well below its
asymptotic cut-off.  \citet{wil05} note that in this limit the
logit-normal distribution approaches the log-normal.  The power-law
normal distribution much more closely captures the smaller high-$N$
tail.  The corresponding cumulative distribution functions (c.d.f.s)
are plotted in Figure~\ref{fig:basalCDF}, where the logit-normal
distribution has been omitted for clarity.  It can be seen, especially
for $R=10^6$, that the log-normal distribution underestimates the
cumulative number of species in both tails, which corresponds to the
skew of the p.d.f., and that even for one hundred realisations the
empirical c.d.f. is far from smooth.  More instructive than the
c.d.f. are the residuals of this plot, that is, the difference between
the instantaneous value of the empirical c.d.f. and the fitting
function.  These are shown in Figure~\ref{fig:basalCDFr} for all three
fitting functions.  The integral of the square of this plot is our
goodness-of-fit measure $k^2$, and the maximum deviation from zero is
the Kolmogorov-Smirnov $d$-measure.  Substantial structure can be seen
in the residuals, especially the central peak for each value of $R$
when examining the power-law normal distribution, which most closely
mimics the tails.  Table~\ref{Tab:Fitting} records the values of $k^2$
and the Kolmogorov-Smirnov $d$ value for each fit, for fitting
parameters minimising $k^2$.  Basal communities are labelled by the
value of resources, $R$, while trophic levels examined in
section~\ref{Sec:Trophic} are labelled according to the trophic level,
$l$.  For the basal food webs the power-law normal fit always
outperforms both the logit-normal and log-normal distributions in
terms of $k^2$, and is only in one case inferior to the logit-normal
distribution as measured by $d$.  A further comparison of the relative
merits of the theoretical distribution functions is given in
section~\ref{Sec:Tails}.

In Figure~\ref{fig:basalFitPar} we plot the dependence of the
parameters of the power-law normal fit on $R$, as well as the two
goodness-of-fit indicators used.  The solid line, marking the
population of the peak of the distribution, indicates the very near
linearity of the value of the peak of the distribution with $\ln R$.
The standard deviation of the distribution increases more rapidly than
linearly, as indicated by the dashed line.  The value of $k^2$
increases with $\ln R$ for two reasons.  Firstly, it is measured on
the full c.d.f. rather than the normalised distribution, and so tends
to increase as the square of the expected number of species, $S$.
Secondly, because it is an integrated measure, it tends to increase
with the width of the distribution, which we characterise by the
standard deviation of the log-normal distribution, $\sigma_{\rm LN}$.
It is more appropriate to use this measure than the standard deviation
of the power-law normal itself since the former corresponds naturally
to the width along the logarithmic population axis.  In
Figure~\ref{fig:basalFitPar} we plot the quantity
\begin{equation}
  K=\frac{1000k^2}{S^2\sigma_{\rm LN}},
  \label{Eq:K}
\end{equation}
which compensates for these effects, and includes a factor of $1000$
to scale it appropriately for that plot.  It can be seen that
intermediate values of $R$ are the best fitted, as measured by either
$K$ or the Kolmogorov-Smirnov $d$, perhaps due to relatively small
amounts of additional structure.

\subsection{Trophic levels}
\label{Sec:Trophic}
\begin{figure}
  \centering
  \includegraphics[width=0.45\textwidth]{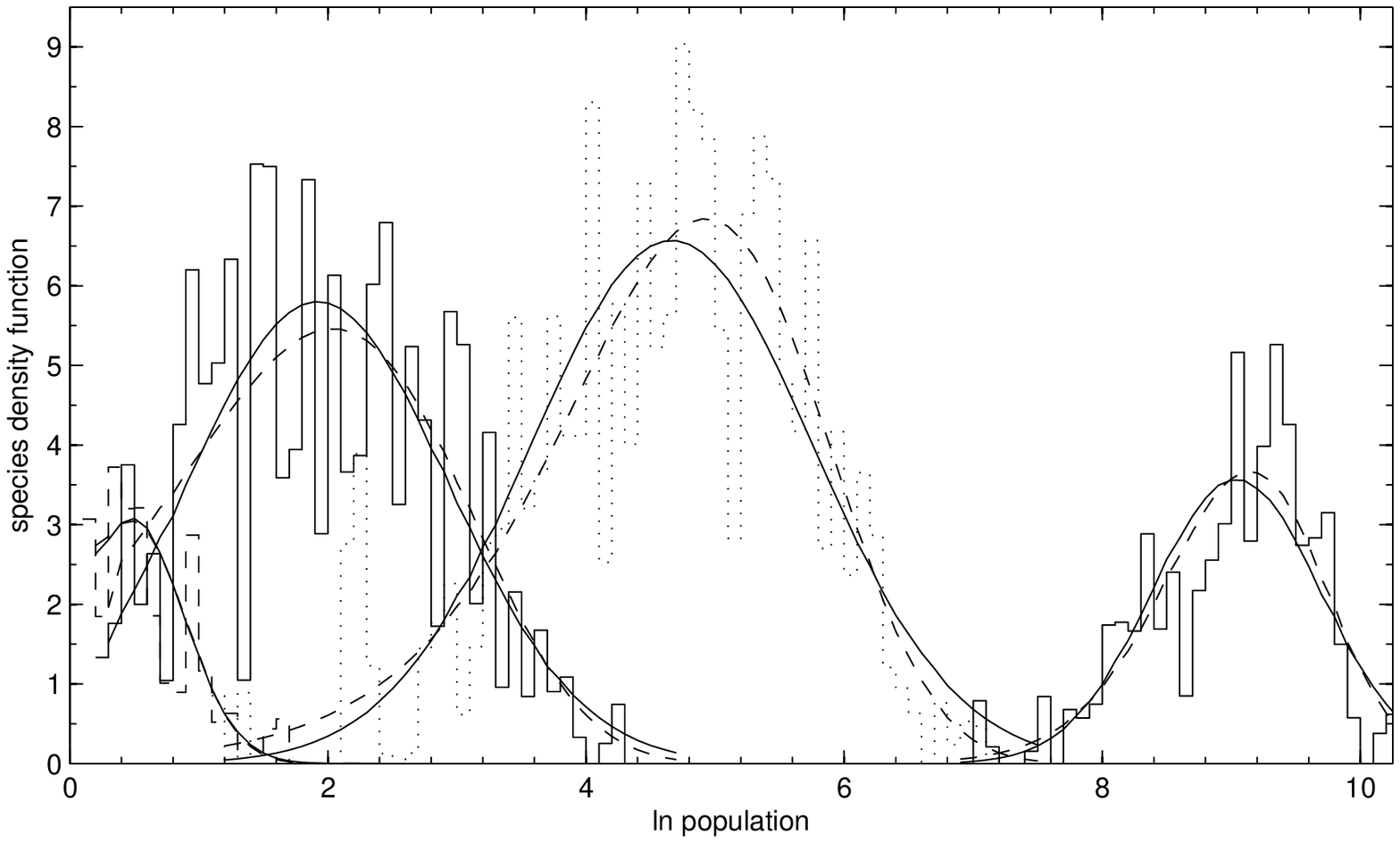}
  \caption{Histograms mark the observed species abundance distribution
    for the four trophic levels found in the ten Webworld communities
    examined.  Trophic levels two and four are marked by dotted and
    dashed lines respectively.  Solid curves mark the optimal
    log-normal fits to each trophic level, and dashed lines the
    optimal power-law normal fits.
  }
  \label{fig:trophicPDF}
\end{figure}
\begin{figure}
  \centering
  \includegraphics[width=0.45\textwidth]{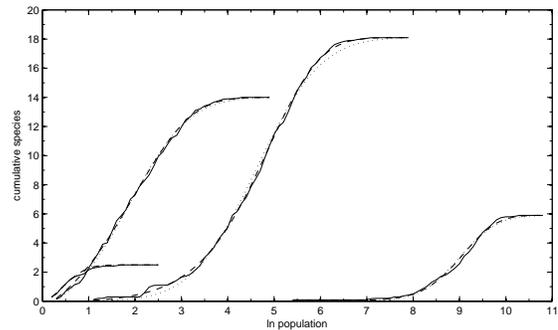}
  \caption{The cumulative species abundance distribution for
    Webworld communities corresponding to the four observed trophic
    levels.  Higher trophic levels are to the left of lower levels,
    having smaller typical populations.  Optimal log-normal fits are
    marked by dotted lines, and optimal power-law normal fits by
    dashed lines.
  }
  \label{fig:trophicCDF}
\end{figure}
Having established that the power-law normal distribution describes
the SAD reasonably well for basal communities, we apply it to
individual trophic levels of full Webworld communities to determine
the relevant fitting parameters.  Due to the small number of food webs
available, and the small number of species in each trophic level for
any one web, it is inappropriate to seek deviations from this
distribution with the data available, although we find that the
power-law normal distribution is adequate, and superior in all cases
to the log-normal distribution, having smaller values of both $k^2$
and $d$.  As indicated by the values given in table~\ref{Tab:Fitting},
the logit-normal distribution marginally improves upon the power-law
normal distribution for trophic levels 1 and 3, but is significantly
inferior to the power-law normal for trophic level 2.  For trophic
level 1, the typical number of species observed per web in the data
examined was only 5.9, the most abundant species being nearly half the
total population of its trophic level.  For trophic level 3 the lower
tail of the distribution was truncated, and although here the
logit-normal distribution performed better than the power-law normal,
it is not clear that the logit-normal is able to adequately reproduce
the whole SAD.  Although four trophic levels were found in the
empirical data, a very small number of species were found in trophic
level 4.  It can be seen in Figure~\ref{fig:trophicPDF} that the
distribution function of this level is little more than the
high-population tail of the distribution function, and no reliable
results can be obtained by its analysis.

\begin{figure}
  \centering
  \includegraphics[width=0.45\textwidth]{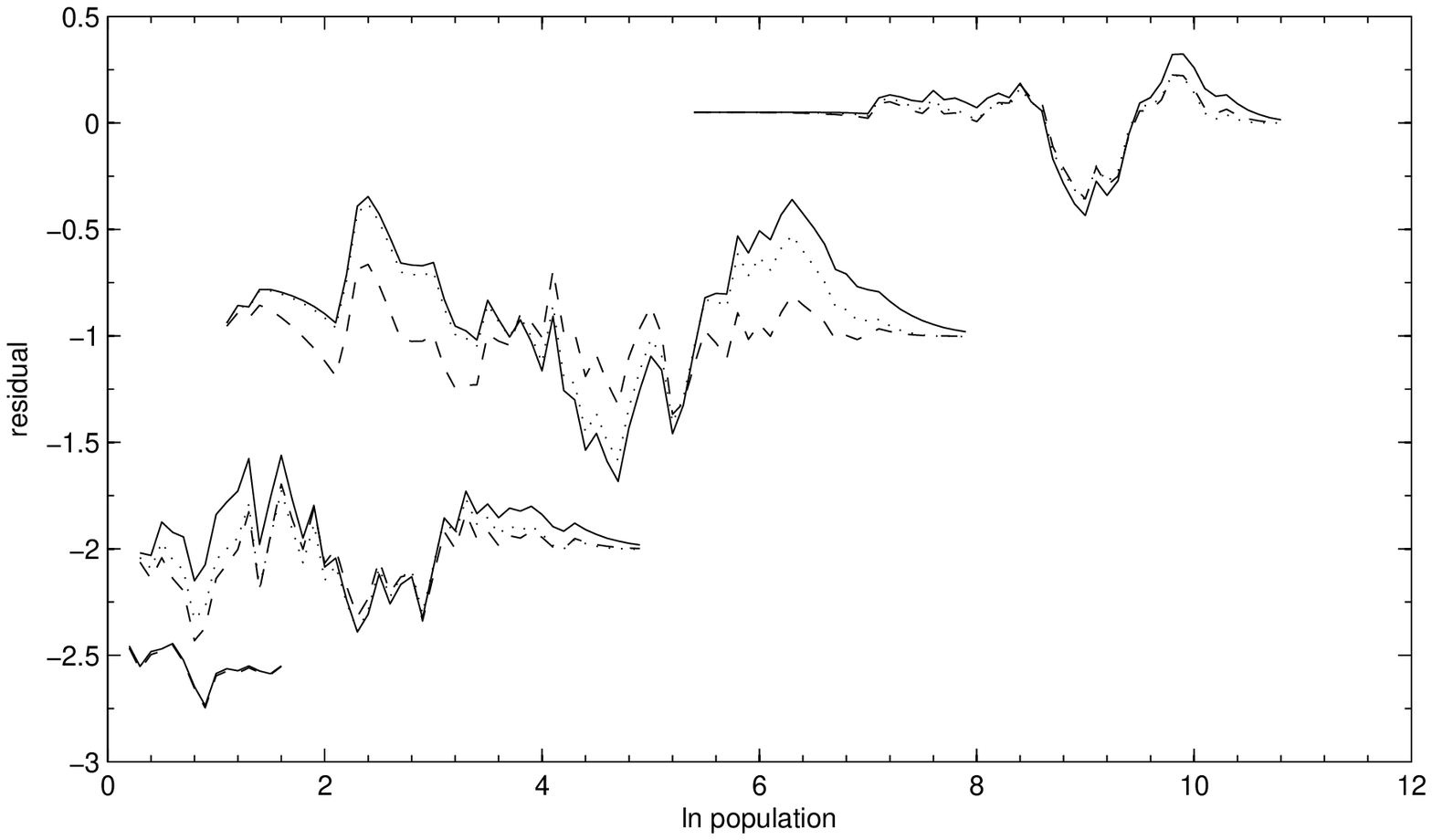}
  \caption{The same data plotted in Figure~\ref{fig:trophicCDF} shown as
    residuals; the solid line corresponds to the empirical c.d.f. minus the
    log-normal distribution, the dotted line to the data minus the
    logit-normal distribution, and the dashed line to the data minus the
    power-law normal distribution.  Offsets of -1.0, -2.0 and -2.5 have been
    applied to data for trophic levels 2, 3 and 4 respectively.  No
    logit-normal fit was obtained for trophic level 4 due to the
    absence of a positive optimal mean.
  }
  \label{fig:trophicCDFr}
\end{figure}

For comprehension of the empirical distribution being fitted we
reproduce, in Figure~\ref{fig:trophicCDF}, the cumulative distribution
function constructed from the simulation data along with the optimal
log-normal and power-law normal fits.  It can be seen clearly from
this figure that the distribution of the second trophic level, which
has the largest number of species in total, is closest in form the
those of the basal communities.  The distribution of trophic level
three, to its left, passes the veil line before a significant fraction
of the low-population tail has been exposed, but is otherwise in good
agreement with the basal community distributions.  The lowest trophic
level, however, seems relatively truncated, resulting in a much
sharper cutoff at large $N$ than is reproduced by either the
log-normal or power-law normal distributions.  The cause of this may
relate to the presence of predators, who can be expected to
preferentially target the most abundant prey, but additional data are
required to investigate this hypothesis.  The residuals of the
c.d.f. fits are shown in Figure~\ref{fig:trophicCDFr}; it is possible
that similar structure in these is present to that seen for the basal
communities in Figure~\ref{fig:basalCDFr}, but the degree of noise is
greater.

\begin{figure}
  \centering
  \includegraphics[width=0.45\textwidth]{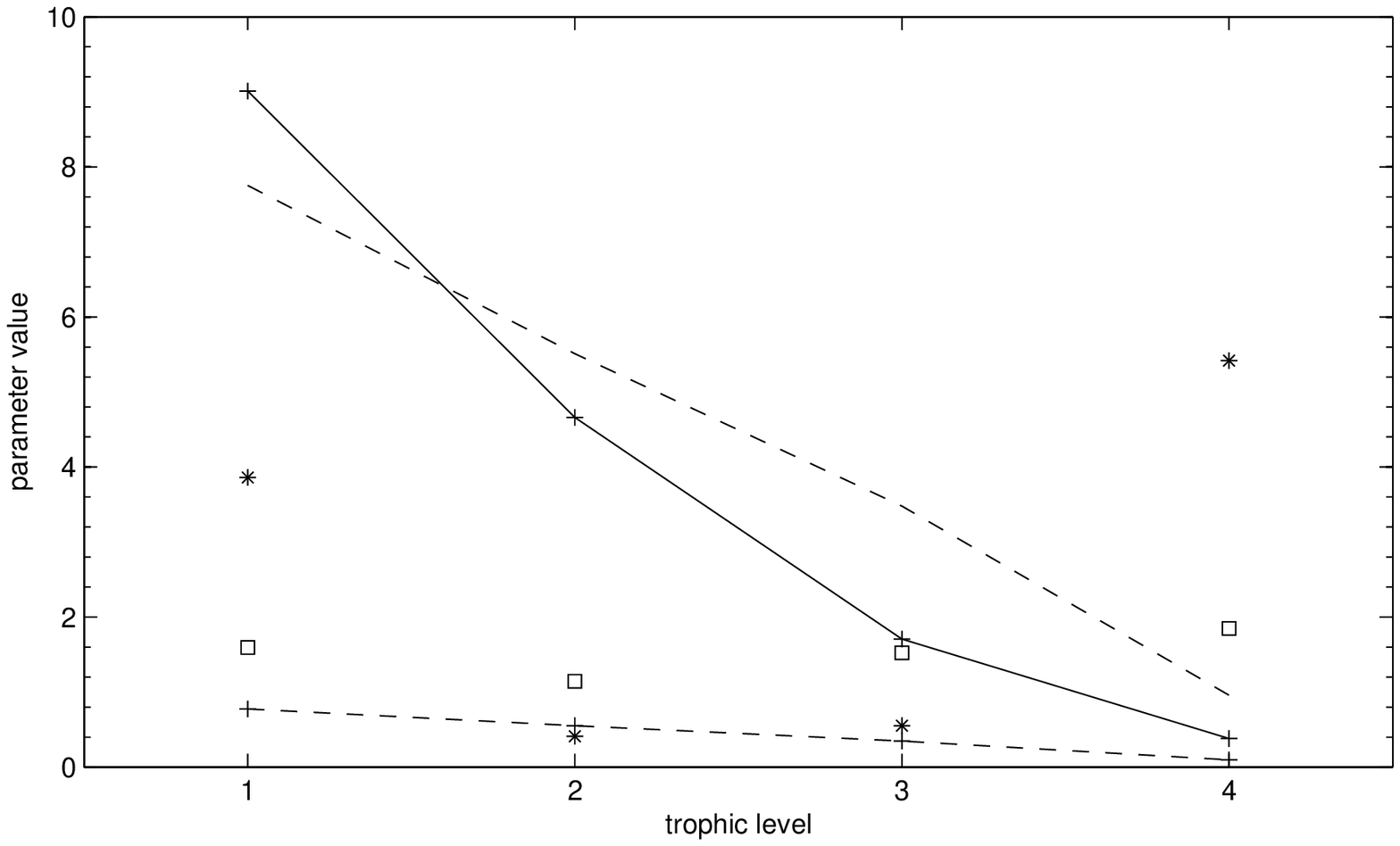}
  \caption{
    Parameters of the power-law normal fit to the trophic community SAD
    for all values of resources examined.  The solid line passes
    through points marking the mean value of $N$.  The lower dashed line 
    marks the standard deviation, while the upper dashed line multiplies
    this quantity by 10 for clarity.  Squares mark the
    Kolmogorov-Smirnov $d$ value.  Stars mark the quantity $K$ defined
    in the text.
  }
  \label{fig:trophicFit}
\end{figure}

In Figure~\ref{fig:trophicFit} the mean and standard deviation of the
power-law normal distribution are plotted as a function of trophic
level.  While the standard deviation appears to decline linearly with
trophic level, the distribution mean may decrease more slowly.
However, if the results for trophic level four are misleading due to
the extremely high position of the veil line, and the distribution of
basal species is possibly altered through predation as discussed, the
reliability of these results is limited.  The quantity $K$, defined in
Eq.~(\ref{Eq:K}), is much better for trophic levels two and three
than for either the basal or fourth trophic level, although only
marginal improvements in the Kolmogorov-Smirnov $d$ value can be seen.

\subsection{Distribution tails}
\label{Sec:Tails}
An advantage of examining computationally derived communities of
species is that extremely large data sets can be constructed with
relative ease, subject only to the availability of computer time.  In
addition, the Webworld model produces complete ecological communities,
and the sampling effects associated with field data are avoidable.  As
such it is much more feasible to examine the form taken by the tails
of the distribution function, which \citet{mcg07} note are subject to
noisy data, but which often contain the main differences between
theoretical distributions.

\begin{figure}
  \centering
  \includegraphics[width=0.45\textwidth]{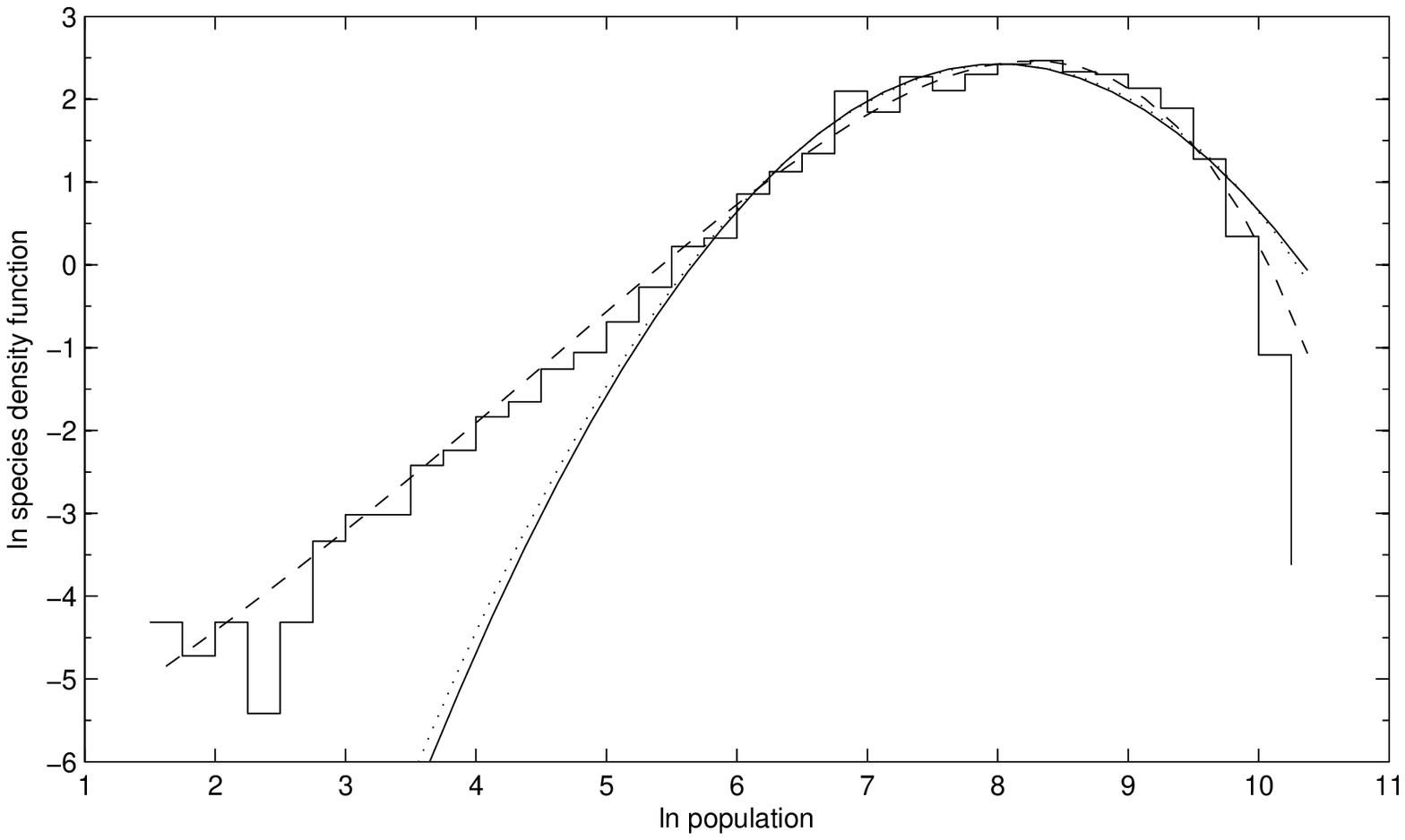}
  \caption{
    The small population tail of the basal community p.d.f. for resources 
    $R=10^6$. The p.d.f. is described in the text.  The solid, dotted and 
    dashed curves mark the log-normal, logit-normal and power-law normal 
    fits to Figure~\ref{fig:basalCDF} respectively.
  }
  \label{fig:lowTail}
\end{figure}

To construct a high-quality empirical SAD whose tails could be
examined, nine hundred simulation runs were performed for the basal
community with $R=10^6$.  The low-population tail of this distribution
is plotted in Figure~\ref{fig:lowTail}, where the logarithm of the binned
species abundance has been taken to expose the tail.  The fact that a
linear regression to this data (not shown) produces a good fit for
$\ln N<7$ implies that in this regime a power-law fit,
\begin{equation}
 P\!\left(\ln N\right)\propto N^a,
\end{equation}
with $a\sim 4/3$, is obeyed.  The power-law normal distribution is
able to reproduce this form reasonably well, while both the log-normal
and logit-normal distributions significantly underestimate the
number of species present.

\begin{figure}
  \centering
  \includegraphics[width=0.45\textwidth]{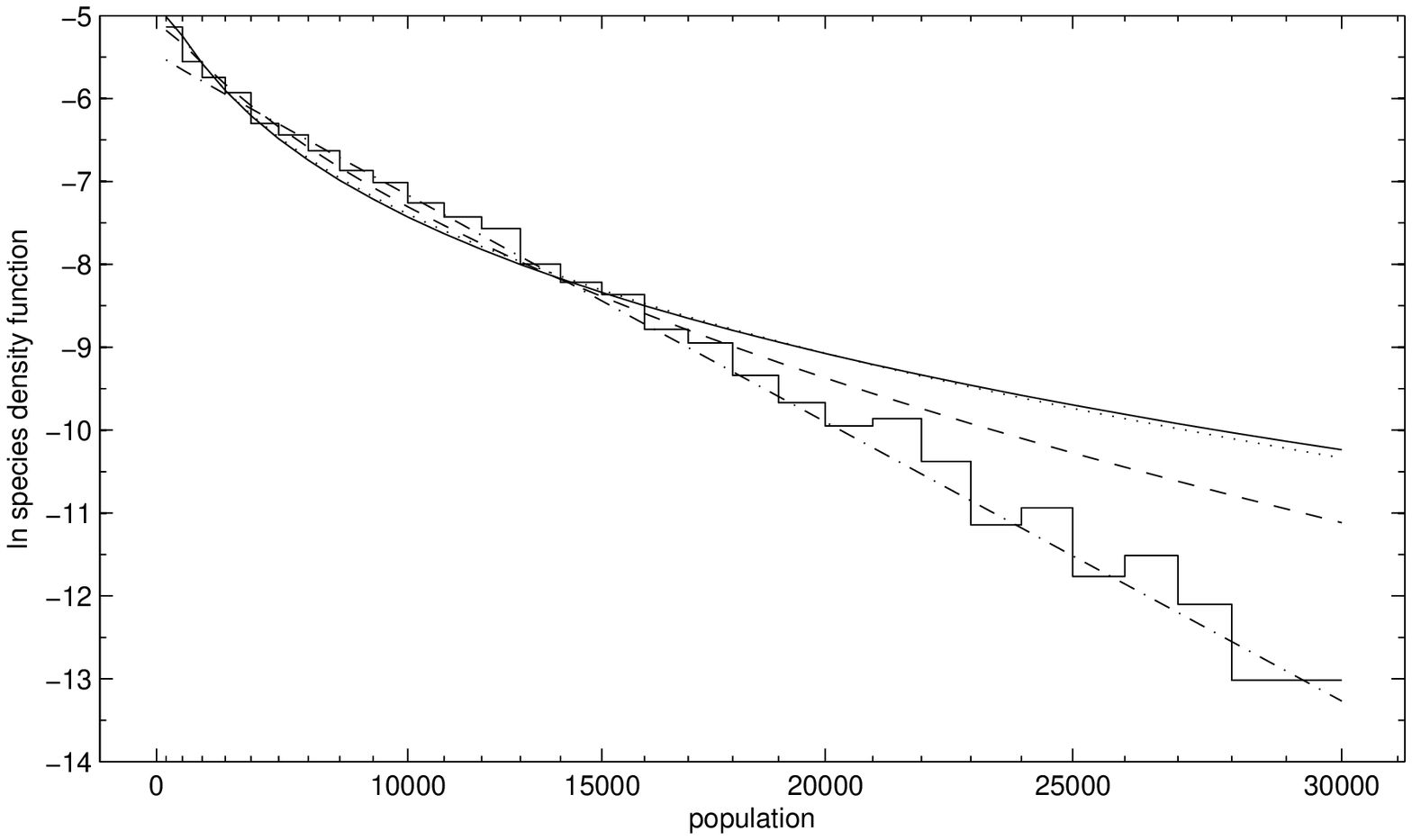}
  \caption{The high population tail of the basal community p.d.f. for
    resources $R=10^6$.  The $x$-axis is linear in $N^{1.4116}$, which
    was found to be the power-law index minimising the $\chi^2$ of the
    regression line, but has been marked with corresponding values of
    $N$ for clarity.  The histogram marks the value of
    $P\!\left(N\right)$, the population density in bins of equal width
    in $N$.  The solid, dotted and dashed curves mark the log-normal,
    logit-normal and power-law normal fits to
    Figure~\ref{fig:basalCDF} respectively.  The dash-dotted line
    indicates the best-fitting regression for $N>2000$.
  }
  \label{fig:highTail}
\end{figure}

The distribution tail for large populations is shown in
Figure~\ref{fig:highTail}.  Here bins have been chosen to be uniform
in width in population, rather than uniform in $\ln N$, in order to
resolve the tail.  The result is that a different version of the
distribution is shown,
\begin{equation}
P\!\left(N\right){\rm d}N=\frac{P\!\left(\ln N\right)}{N}{\rm d}N,
\end{equation}
which, when integrated with respect to $N$, gives the c.d.f.  Note
that in order to highlight the form of the decay, the population axis
has been stretched to a power-law.  The regression line, plotted as a
dash-dot line, indicates that the high-population tail has the form
\begin{equation}
P\!\left(N\right){\rm d}N\propto
\exp\left\{-\left(\frac{N}{7140}\right)^{1.4116}\right\}{\rm d}N.
\end{equation}
As can be seen in Figure~\ref{fig:highTail}, this form of the decay
declines more rapidly with $N$ than any of the log-normal,
logit-normal or power-law normal distributions examined.

\begin{figure}
  \centering
  \includegraphics[width=0.45\textwidth]{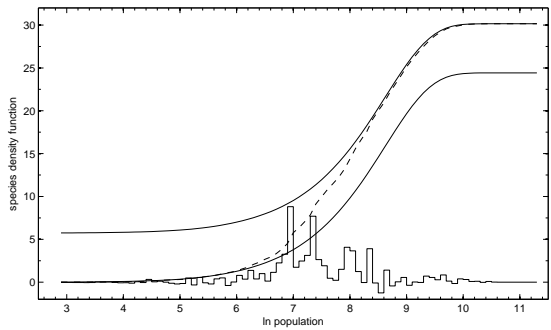}
  \caption{
    The c.d.f. for the basal community with resources $R=10^6$, as shown
    in Figure~\ref{fig:basalCDF}, is shown as a dashed line.  The
    solid lines mark the c.d.f. constructed from fits to the tails as
    described in the text.  The histogram marks the residuals of the
    p.d.f. of this fit.
  }
  \label{fig:tails}
\end{figure}

Having established probable forms for the low- and high-population
tails by regression to Figures~\ref{fig:lowTail} and
\ref{fig:highTail}, we combine these into a distribution which has the
minimum value of the two tail-fitting functions for all $N$.  In
addition to the dashed line marking the empirical c.d.f., identical to
that shown in Figure~\ref{fig:basalCDF}, this fit is shown in
Figure~\ref{fig:tails} in two forms.  The lower plot is the
c.d.f. integrated from zero species at $N=0$, and the upper curve is
integrated down from the observed number of species so as to converge
with the empirical distribution at large $N$.  The fact that the
latter curve is above the former indicates that the combined
distribution underestimates the total number of species, implying that
it under-predicts the p.d.f. near the peak, to which it was not
fitted.  Figure~\ref{fig:tails} therefore also plots the residuals of
the tail-fitting distribution as a histogram.  There appear to be at
least three peaks in the residuals, making it difficult to identify a
plausible general form.  Since we do not have unrelated basal food
webs to examine, in particular to establish what parameters of the
tail distributions are generic and whether the residuals show a common
pattern, it is not appropriate to draw further conclusions about the
central part of the distribution.  We are also unable to ascribe a
goodness-of-fit to the tail-based distribution due to its inability to
reproduce the peak of the distribution.

\section{Conclusions}
\label{Sec:Conclusions}
We have investigated the form of the species abundance distribution
empirically derived from simulation results of the Webworld food web
model.  This model was created to examine patterns of food web
assembly, and the form of the species abundance distribution (SAD) was
not a factor in its construction.  Rather, the use of population
dynamics to establish the success of particular species and feeding
strategies within the community lead naturally to variation in the
abundance of species which appears similar to the SADs identified from
real ecosystems.  By investigating the empirical SAD from the
simulations in the same manner as data from real ecosystems we are
able to characterise not only the peak of the distribution, which is
frequently observed to have a form similar to the log-normal
distribution, but to examine in detail parts of the distribution
difficult to obtain data on from field studies.  We agree with the
conclusion of \citet{wil05} that the logit-normal distribution fits
better, but with particular reference to the tails of the distribution
find that the power-law normal distribution function is better still.
In particular, the log-normal and logit-normal distributions predict
that the number of species with population $N$ falls more rapidly with
decreasing $N$ than we obtain from our simulation results, which the
power-law normal distribution matches very well in this tail.

The presence of structure in Figure~\ref{fig:basalCDFr} suggests that
a more complicated function is needed to properly reproduce the
observed SAD, but we have not been able to examine the reproducibility
of this remaining structure.  All the food webs examined were created
for the same set of possible features and the same environment
species.  To fully explore the results even for a single value of $R$
would require the use of food webs constructed for `worlds' with
different environment species and feature sets.  In undertaking such a
programme it would first be necessary to establish whether such
parameters as the mean and variance of the fitted distribution
changed, or more generally to construct the meta-distribution of a
large number of Webworld `worlds' and test, using the
Kolmogorov-Smirnov $d$ value, whether the empirical distribution
constructed from webs of a single family was consistent with the
meta-distribution.

We find that the power-law normal distribution identified as well
describing the SAD of a basal community is also successful in
describing individual trophic levels of a food web.  It is
particularly descriptive of the second trophic level, which can be
seen in Figures~\ref{fig:trophicPDF} and \ref{fig:trophicCDF} to be
the most completely realised by our empirical data.  The higher
trophic levels can also be expected to be well-fitted by the power-law
normal distribution, although the truncation of the distribution at
low populations results in the log-normal and logit-normal
descriptions also being adequate.  The empirical distribution of the
lowest trophic level is more sharply truncated at high populations
than seen for other communities, the reason for which would require
substantial additional investigation.  Unlike the case of examining
basal communities at different values of $R$, only a small number of
trophic levels are ever possible, and hence the relation between them
is harder to quantify.  While it would be possible to construct
meta-distributions from larger numbers of food webs, it is more
feasible to first examine the agreement between the meta-distributions
of basal communities and the constituent distributions.  If there is
good agreement, the agreement between the meta-distribution and the
trophic distributions should be examined.  If not, then a very large
number of communities need to be evolved in the same environment in
order to study the relation between trophic levels, potentially also
examining the effect of different values of $R$.  The main problem in
investigating the SAD of numerically modelled ecosystems is the
extensive computer time required to provide data.

The SADs constructed for this paper are complete not only in the sense
that they contain all individuals present in the sample area, but also
in that they do not feature immigrant or transient species, which can
contribute to the low-population tail without representing a viable
population.  While features such as immigration from surrounding
communities can easily be incorporated into our model, as can finite
population effects, their exclusion demonstrates the existence of an
extensive low-population tail to the distribution even for a closed
ecosystem.  This contrasts with the proposal by \citet{mag03} that the
low-population tail is a log-series distribution of ``occasional''
species added to a core log-normal distribution.  Although we do not
agree with \citet{mcg03a} that left-skew is purely an effect of
sampling, it may be the case that the left-skew of incomplete
samples does not reflect the underlying distribution.

\citet{mcg07} observe that most proposed SADs are similar to one
another except in the tails, which is precisely the region which field
observations are least able to address due to paucity of data.  This
issue can be addressed by the use of any model which can produce
multiple independent realisations of its dynamics from which a
composite SAD can be constructed, but this process can only be used to
inform the analyses which should be performed on ecological data,
since it is not known a priori that any given model accurately
reproduces the real SAD.  A virtue of the Webworld model is that is
produces a plausible SAD without any such consideration having been
used during the model design, being based rather on plausible
ecological rules.

\section*{Acknowledgements}
The authors thank Carlos A. Lugo for providing additional simulation
data.  This work was supported by EPSRC under grant GR/T11784.

\newpage

\begin{table}
\caption{Comparison of measures of goodness-of-fit for the log-normal, 
  logit-normal and power-law normal distributions to basal community SADs 
  and trophic SADs.
}
\centering
\label{Tab:Fitting}
\begin{tabular}{ccccc}
\hline\noalign{\smallskip}
Community & Species & \multicolumn{3}{c}{$k^2$}   \\[3pt]
          & $S$     & Log    & Logit  & Power-law \\[3pt]
$R=10^3$  & 11.22   & 0.0599 & 0.0390 & 0.0314    \\
$R=10^4$  & 18.15   & 0.2333 & 0.1571 & 0.1175    \\
$R=10^5$  & 25.42   & 0.8820 & 0.6296 & 0.1676    \\
$R=10^6$  & 30.16   & 1.5759 & 1.1752 & 0.4708    \\
$l=1$     &  5.9    & 0.1532 & 0.0902 & 0.0877    \\
$l=2$     & 18.1    & 0.7972 & 0.5328 & 0.1480    \\
$l=3$     & 14.0    & 0.2111 & 0.1126 & 0.1093    \\
\hline\noalign{\smallskip}
Community & Species & \multicolumn{3}{c}{Kolmogorov-Smirnov $d$} \\[3pt]
          & $S$     & Log    & Logit  & Power-law            \\[3pt]
$R=10^3$  & 11.22   & 1.0648 & 1.1094 & 0.9990               \\
$R=10^4$  & 18.15   & 1.3244 & 1.0409 & 1.1660               \\
$R=10^5$  & 25.42   & 1.4826 & 1.3669 & 0.9919               \\
$R=10^6$  & 30.16   & 2.0255 & 1.7661 & 1.6167               \\
$l=1$     &  5.9    & 1.8391 & 1.5798 & 1.5934               \\
$l=2$     & 18.1    & 2.0051 & 1.7753 & 1.1430               \\
$l=3$     & 14.0    & 1.8270 & 1.4797 & 1.5241               \\
\hline\noalign{\smallskip}
\end{tabular}
\end{table}


\begin{thebibliography}{21}
\providecommand{\natexlab}[1]{#1}

\bibitem[{{Fisher} \textit{et~al.}(1943){Fisher}, {Corbet}, \&
  {Williams}}]{fis43}
\textsc{{Fisher}, R.~A., {Corbet}, A.~S., {Williams}, C.}, 1943; \textit{The
  relationship between the number of species and the number of individuals in a
  random sample from an animal population}.
\newblock J. Animal Ecology \textbf{12} 42

\bibitem[{{Preston}(1948)}]{pre48}
\textsc{{Preston}, F.~W.}, 1948; \textit{The commonness and rarity of species}.
\newblock Ecology \textbf{29} 254

\bibitem[{{Gray}(1987)}]{gra87}
\textsc{{Gray}, J.~S.}, 1987; \textit{Species-abundance patterns}.
\newblock In {O}rganization of {C}ommunities {P}ast and {P}resent (J.~H.~R.
  {Gee}, P.~S. {Giller}, eds.) (Blackwell Science, Oxford), 53--68

\bibitem[{{Marquet} \textit{et~al.}(2003){Marquet}, {Fern\'andez}, \&
  {Cofre}}]{mar03}
\textsc{{Marquet}, P.~A., {Fern\'andez}, J.~A., {Cofre}, H.}, 2003;
  \textit{Breaking the stick in space: of niche models, metacommunities and
  patterns in the relative abundance of species}.
\newblock In {M}acroecology: {C}oncepts and {C}onsequences (T.~M. {Blackburn},
  K.~J. {Gaston}, eds.) (Blackwell Science, Oxford), 64--86

\bibitem[{{May}(1975)}]{may75}
\textsc{{May}, R.~M.}, 1975; \textit{Patterns of species abundance and
  diversity}.
\newblock In {Ecology and evolution of communities} (M.~L. {Cody}, J.~M.
  {Diamond}, eds.) (Belknap Press, Harvard), 81--120

\bibitem[{{McGill} \textit{et~al.}(2007){McGill}, {Etienne}, {Gray}, {Alonso},
  {Anderson}, {Benecha}, {Dornelas}, {Enquist}, {Green}, {He}, {Hurlbert},
  {Magurran}, {Marquet}, {Maurer}, {Ostling}, {Soykan}, {Ugland}, \&
  {White}}]{mcg07}
\textsc{{McGill}, B.~J., {Etienne}, R.~S., {Gray}, J.~S., {Alonso}, D.,
  {Anderson}, M.~J., {Benecha}, H.~K., {Dornelas}, M., {Enquist}, B.~J.,
  {Green}, J.~L., {He}, F., {Hurlbert}, A.~H., {Magurran}, A.~E., {Marquet},
  P.~A., {Maurer}, B.~A., {Ostling}, A., {Soykan}, C.~U., {Ugland}, K.~I.,
  {White}, E.~P.}, 2007; \textit{Species abundance distributions: moving beyond
  single prediction theories to integration within an ecological framework}.
\newblock Ecology Letters \textbf{10} 995

\bibitem[{{Whittaker}(1965)}]{whi65}
\textsc{{Whittaker}, R.~H.}, 1965; \textit{Dominance and diversity in land
  plant communities}.
\newblock Science \textbf{147} 250

\bibitem[{{Drossel} \textit{et~al.}(2001){Drossel}, {Higgs}, \&
  {McKane}}]{dro01}
\textsc{{Drossel}, B., {Higgs}, P.~G., {McKane}, A.~J.}, 2001; \textit{The
  influence of predator-prey population dynamics on the long term evolution of
  food web structure}.
\newblock J. Theor. Biol. \textbf{208} 91

\bibitem[{{Caldarelli} \textit{et~al.}(1998){Caldarelli}, {Higgs}, \&
  {McKane}}]{cal98}
\textsc{{Caldarelli}, G., {Higgs}, P.~G., {McKane}, A.~J.}, 1998;
  \textit{Modelling coevolution in multispecies communities}.
\newblock J. Theor. Biol. \textbf{193} 345

\bibitem[{{Drossel} \textit{et~al.}(2004){Drossel}, {McKane}, \&
  {Quince}}]{dro04}
\textsc{{Drossel}, B., {McKane}, A.~J., {Quince}, C.}, 2004; \textit{The impact
  of nonlinear functional responses on the long-term evolution of food web
  structure}.
\newblock J. Theor. Biol. \textbf{229} 539

\bibitem[{{Quince} \textit{et~al.}(2005){Quince}, {Higgs}, \& {McKane}}]{qui05}
\textsc{{Quince}, C., {Higgs}, P.~G., {McKane}, A.~J.}, 2005;
  \textit{Topological structure and interaction strengths in model food webs}.
\newblock Ecol. Model. \textbf{187} 389

\bibitem[{{Drossel} \& {McKane}(2003)}]{dro03}
\textsc{{Drossel}, B., {McKane}, A.~J.}, 2003; \textit{Modelling food webs}.
\newblock In Handbook of {G}raphs and {N}etworks ({S. Bornholdt and H.G.
  Schuster}, ed.) (Wiley-VCH), 218--247

\bibitem[{{McKane} \& {Drossel}(2005)}]{mck05}
\textsc{{McKane}, A.~J., {Drossel}, B.}, 2005; \textit{Modelling evolving food
  webs}.
\newblock In {D}ynamical {F}ood {W}ebs (P.~C. de~Ruiter, V.~Wolters, J.~C.
  Moore, eds.) (Elsevier, Singapore), 74--88

\bibitem[{{McKane} \& {Drossel}(2006)}]{mck06}
\textsc{{McKane}, A.~J., {Drossel}, B.}, 2006; \textit{Models of food web
  evolution}.
\newblock In {E}cological {N}etworks: {L}inking {S}tructure to {D}ynamics in
  {F}ood {W}ebs (Oxford University Press), 223--243

\bibitem[{{Bain} \& {Engelhardt}(1991)}]{bai91}
\textsc{{Bain}, L.~J., {Engelhardt}, M.}, 1991; Introduction to {P}robability
  and {M}athematical {S}tatistics (Duxbury)

\bibitem[{{Hayter}(2002)}]{hay02}
\textsc{{Hayter}, A.~J.}, 2002; Probability and {S}tatistics for {E}ngineers
  and {S}cientists (Duxbury), 2nd edn.

\bibitem[{{McGill}(2003{\natexlab{\textit{a}}})}]{mcg03}
\textsc{{McGill}, B.~J.}, 2003{\natexlab{\textit{a}}}; \textit{Strong and weak
  tests of macroecological theory}.
\newblock Oikos \textbf{102} 679

\bibitem[{{Williamson} \& {Gaston}(2005)}]{wil05}
\textsc{{Williamson}, M., {Gaston}, K.~J.}, 2005; \textit{The lognormal
  distribution is not an appropriate null hypothesis for the species adundance
  distribution}.
\newblock J. Animal Ecology \textbf{74} 409

\bibitem[{{Magurran}(1988)}]{mag88}
\textsc{{Magurran}, A.~E.}, 1988; Ecological diversity and its measurement
  (Cambridge University Press)

\bibitem[{{Magurran}(2003)}]{mag03}
\textsc{{Magurran}, A.~E.}, 2003; \textit{Explaining the excess of rare species
  in natural species abundance distributions}.
\newblock Nature \textbf{422} 714

\bibitem[{{McGill}(2003{\natexlab{\textit{b}}})}]{mcg03a}
\textsc{{McGill}, B.~J.}, 2003{\natexlab{\textit{b}}}; \textit{Does Mother
  Nature really prefer rare species or are log-left-skewed SADs a sampling
  artefact?}
\newblock Ecology Letters \textbf{6} 766

\end{thebibliography}
\end{document}